\documentclass[prl,aps,showpacs,floatfix,nofootinbib,twocolumn,letterpaper]{revtex4}
\usepackage{epsfig}
\usepackage{graphicx}
\usepackage{amssymb}
\usepackage{amsmath}
\usepackage{amsfonts}
\newcommand{\lb}{\langle}
\newcommand{\rb}{\rangle}
\newcommand{\be}{\begin{equation}}
\newcommand{\ee}{\end{equation}}
\newcommand{\ba}{\begin{eqnarray}}
\newcommand{\ea}{\end{eqnarray}}
\newcommand{\ad}{a^\dagger}
\newcommand{\Tr}{{\rm Tr}}
\newcommand{\ve}{\varepsilon}
\newcommand{\bi}{\bibitem}
\newcommand{\ma}{\mathcal{A}}
\begin{document}
\title{Odd-particle systems in the shell model Monte Carlo: circumventing a sign problem}
\author{Abhishek Mukherjee and Y. Alhassid}
\affiliation{Center for Theoretical Physics, Sloane Physics
Laboratory, Yale University, New Haven, CT 06520}
\date{\today}
\begin{abstract}
We introduce a novel method within the shell model Monte Carlo approach to calculate the ground-state energy of a finite-size system with an odd number of particles by using the asymptotic behavior of the imaginary-time single-particle Green's functions. The method circumvents the sign problem that originates from the projection on an odd number of particles and has hampered direct application of the shell model Monte Carlo method to odd-particle systems. We apply this method to calculate pairing gaps of nuclei in the iron region. Our results are in good agreement with experimental pairing gaps.
\end{abstract}
\pacs{21.60.Ka, 21.60.Cs, 21.60De, 21.10.Dr, 27.40.+z, 27.40.+e, 26.50.+x}
\maketitle
\emph{Introduction.} The shell model Monte Carlo (SMMC) approach~\cite{lan93,alh94,koo97,alh01} has been used successfully to calculate statistical properties of nuclei~\cite{nak97,alh99,alh08} within the framework of the configuration-interaction shell model. Recently, this method has also been applied to trapped cold atom systems~\cite{zin09,gil11}. The SMMC method enables calculations in model spaces that are many orders of magnitude larger than those that can be treated by conventional diagonalization methods.

For typical effective nuclear interactions, the SMMC method breaks down at low temperatures because of the so-called fermionic sign problem, leading to large statistical errors. In the grand-canonical ensemble the sign problem can be avoided by constructing good-sign interactions that include the dominant collective components of effective nuclear interactions~\cite{duf96}. The remaining part of the effective
interaction can be accounted for by using the method of Ref.~\onlinecite{alh94}.

In finite-size systems, such as nuclei, it is necessary to use the canonical ensemble, in which the number of particles is fixed.  This particle-number projection gives rise to an additional sign problem when the number of particles is odd, leading to a rapid growth of statistical errors at low temperatures even for good-sign interactions. Consequently, it has been a major challenge to make accurate estimates for the ground-state energy of odd-particle systems in SMMC. Accurate ground-state energies are necessary for the calculation of level densities and pairing gaps (i.e., odd-even staggering of binding energies).

Here we develop a method based on the asymptotic behavior of the imaginary-time single-particle Green's functions of an even-particle system to calculate ground-state energies of neighboring odd-particle systems. This method is somewhat similar in spirit to a technique used in lattice quantum chromodynamics to extract hadron masses (see, e.g., in Ref.~\onlinecite{gup97}). We apply our Green's function method to calculate pairing gaps of nuclei in the iron region using the complete $fp+g_{9/2}$ shell model space.

\emph{Green's functions in SMMC.} The SMMC method is based on the Hubbard-Stratonovich representation of the imaginary-time propagator, $e^{-\beta H} = \int D[\sigma]G(\sigma)U_{\sigma}(\beta)$,
where $\beta$ is the inverse temperature, $H$ is the Hamiltonian, $D[\sigma]$ is the integration measure, $G(\sigma)$ is a Gaussian weight, and $U_{\sigma}(\beta)$ is the propagator of non-interacting nucleons moving in external auxiliary fields $\sigma$ that depend on the imaginary time $\tau$ ($0\leq\tau\leq\beta$). The canonical thermal expectation value of
an observable $\hat O$ is given by $\lb \hat O \rb = \int D[\sigma] G(\sigma )\Tr_{\ma} [\hat O U_{\sigma}(\beta) ]/ \int D[\sigma]G(\sigma)\Tr_{\ma} U_{\sigma}(\beta)$,
where $\Tr_{\ma}$ denotes a trace over the subspace of a fixed number of particles $\ma$. In actual calculations we project on both proton number $Z$ and neutron number $N$, and in the following $\ma$ will denote $(Z, N)$.

For a quantity $X_\sigma$ that depends on the auxiliary fields $\sigma$, we define
\be
\label{eqexpt}
\overline{X}_{\sigma} \equiv \frac {\int D[\sigma ] |W(\sigma)| X_{\sigma} \Phi_{\sigma}} { \int D[\sigma] |W(\sigma)| \Phi_{\sigma}},
\ee
where $W(\sigma) = G(\sigma) \Tr_{\ma} U_{\sigma}$
and $\Phi_{\sigma}=W(\sigma)/|W(\sigma)|$ is the sign. With this definition, the above thermal expectation of an observable $\hat O$ can be written as  $\lb \hat O \rb = \overline{\lb \hat O \rb}_{\sigma}$, where $\lb \hat O \rb_{\sigma}={\Tr_{\ma} [\hat O U_{\sigma}(\beta) ]/\Tr_{\ma} U_{\sigma} (\beta)}$.
 In SMMC we choose $M$ samples $\sigma_k$ according to the weight function $|W(\sigma)|$,
and estimate the average quantity in (\ref{eqexpt}) by $\overline{X}_{\sigma} \approx \sum_k  X_{\sigma_k} \Phi_{\sigma_k}/ \sum_k \Phi_{\sigma_k}$.

For an even number of particles with a good-sign interaction, the average value of the sign $\Phi_{\sigma}$ remains close to 1.
However, when the number of particles is odd, the average sign decays towards zero as the temperature is lowered. This leads to rapidly growing errors, hampering the direct application of SMMC at low temperatures for odd-particle systems.

For a rotationally invariant and time-independent Hamiltonian, we define the following scalar imaginary-time Green's functions~\cite{G-tensor}
\be\label{green}
G_{\nu}(\tau) = \frac{\Tr_{\ma}\left[~e^{-\beta H} \mathcal{T} \sum_m a_{\nu m}(\tau) \ad_{\nu m}(0)\right] }{\Tr_{\ma}~e^{-\beta H}},
\ee
where $\nu \equiv (n l j)$ labels the nucleon single-particle orbital with radial quantum number $n$, orbital angular momentum $l$ and total spin $j$. Here $\mathcal{T}$ denotes time ordering and $a_{\nu m}(\tau)\equiv e^{\tau H} a_{\nu m} e^{-\tau H}$ is an annihilation operator of a nucleon at imaginary time $\tau$ ($-\beta \leq \tau\leq \beta$) in a single-particle state with orbital $\nu$ and magnetic quantum number $m$ ($-j\leq m \leq j$).

Using the Hubbard-Stratonovich transformation, the Green's functions defined in (\ref{green}) can be written in a form suitable for SMMC calculations
\be
G_{\nu}(\tau) = \left \{ \begin{array}{ll}
                \overline{\displaystyle \sum_{m}\left [{\bf U}_{\sigma}(\tau) ({\bf I}- \lb\hat\rho \rb_{\sigma} \right ]_{\nu m, \nu m}} & \mbox{ for }\tau>0\\
                                    &   \\
                \overline{\displaystyle \sum_{m} \left [ \lb \hat\rho \rb_{\sigma} {\bf U}^{-1}_{\sigma}(|\tau|) \right ]_{\nu m, \nu m}} & \mbox{ for }\tau \leq 0
                \end{array} \right . \;,
\ee
where we have used the notation in Eq.~(\ref{eqexpt}). Here ${\bf U}_\sigma(\tau)$ and ${\bf I}$ are matrices in the single-particle space representing the propagator $U_\sigma(\tau)$ and the identity, respectively. $\lb \hat\rho \rb_{\sigma}$ is a matrix in the single-particle space whose $\nu m,\nu' m'$ matrix element $\lb \hat\rho_{\nu m,\nu' m'} \rb_{\sigma}$ is defined in terms of the one-body density operator   $\hat\rho_{\nu m,\nu'm'} = \ad_{\nu' m'} a_{\nu m}$.

 Assuming $\ma$ is an even-even nucleus, $\ma_{\pm} \equiv (Z,N\pm1)$ are neighboring odd-even nuclei with odd number of neutrons.
We denote by $J n$ ($n=0,1,2,\ldots$) the $n$-th excited state with total spin $J$ and define the energy differences
$\Delta E_J(\ma_\pm) = E_{J 0}(\ma_{\pm})-E_{00}(\ma)$, where $E_{J n}(\ma)$ is the energy of the state $J n$ with  particle number $\ma$. Assuming that the ground state of the even-even nucleus has spin zero, $\Delta E_J(\ma_\pm)$ is the energy difference between the lowest state of a given spin $J$ in the odd-even nucleus $\ma_\pm$ and the ground state of the even-even nucleus $\ma$. Assuming that the ground state of the odd-even nucleus $\ma_{\pm}$ is $J=j$, where $j$ is one of the single-particle orbital spin values, its corresponding energy is given by
$E_{\rm gs}(\ma_{\pm}) = E_{00}(\ma) + \Delta E_{\rm min}(\ma_{\pm})$,
where $\Delta E_{\rm min}$ is the minimum of $\Delta E_{j}(\ma_\pm)$ over the possible values of $j$.

The neutron Green's function $G_{\nu}(\tau)$ that corresponds to an orbital with angular momentum $j$  can be written as
\begin{widetext}
\be\label{G-expansion}
G_{\nu}(\tau) = C(\beta) e^{- \Delta E_j(\ma_{\pm}) |\tau|}
      \left [ 1 + \sum_{\substack {J n \neq 0 0 \\ J' n' \neq j0}} R_{J n}^{ J' n'}(\ma_{\pm},\nu)  e^{-|\tau|[ E_{J' n'}(\ma_{\pm})-E_{j0}(\ma_{\pm})]}
      e^{-(\beta-|\tau|)[E_{J n}(\ma) - E_{00}(\ma)] } \right ]
\ee
\end{widetext}
where the $+$ ($-$) subscript should be used for $\tau > 0$ ($\tau \leq 0$) and $C(\beta)$ is a $\tau$-independent constant.
 $R_{J n}^{J' n'}(\ma_\pm,\nu)$ are scaled weights defined by
$R_{J n}^{J' n'}(\ma_{+},\nu)= |( J' n' ||\ad_{\nu}|| J n )|^2/ |( j 0 ||\ad_{\nu}|| 00 )|^2$
and $R_{J n}^{J' n'}(\ma_{-};\nu)= |(J' n' ||a_{\nu}|| J n )|^2/|( j 0 ||a_{\nu}|| 00 )|^2$, where $( J' n' ||\ad_\nu|| J n)$  and $( J' n' ||a_\nu|| J n )$ are reduced matrix elements of $\ad_\nu$ and $a_\nu$
between the state $J n$ in $\ma$ and the states $J' n'$ in $\ma_+$ and $\ma_-$, respectively.

When all terms in the summation on the r.h.s. of Eq. (\ref{G-expansion}) are small, the Green's function can be well
approximated by a single exponential, $G_{\nu}(\tau) \sim  e^{- \Delta E_j(\ma_{\pm}) |\tau|}$. In this asymptotic regime for $\tau$, we can calculate
$\Delta E_j(\ma_{\pm})$, and hence $E_{\rm gs}(\ma_{\pm})$ from the slope of $\ln G_{\nu}(\tau)$. This is the method we use here
to calculate the ground-state energy of odd-A nuclei with odd number of neutrons. The ground-state energy of odd-A nuclei with odd number of protons can be similarly calculated using the proton Green's functions.

In principle, the asymptotic regime is accessed in the limit $\beta \to \infty$. However, in a shell-model Hamiltonian with discrete, well separated energy levels, only a few transitions give significant contributions. If the relative contribution from the sum in Eq.~(\ref{G-expansion}) is less than a few percent, then (assuming that $|\tau| \sim 1 $ MeV) the sensitivity of
the slope of $\ln G_\nu(\tau)$ to this contribution is about a few tens of keV, which is comparable to our target accuracy. For low- and medium-mass nuclei,
we expect the energy differences to be $\gtrsim 1$ MeV and the scaled weights to be much smaller than one. Thus, calculations with $\beta$ of a few MeV$^{-1}$ and with an asymptotic regime of $\tau \sim 1$ MeV should be sufficient. This can be validated explicitly in $sd$-shell nuclei (see below), whose Hamiltonian can be diagonalized numerically. For larger model spaces, it is not possible to calculate explicitly the corrections in the sum of Eq.~(\ref{G-expansion}), and the asymptotic region has to be determined by the goodness of the linear fits to $\ln G_{\nu}(\tau)$.

\emph{Results.}
We first tested the Green's function method in $sd$-shell nuclei and then applied it to medium-mass nuclei in the complete ($pf + g_{9/2}$) shell. In these nuclei, we carried out calculations for several values of $\beta$ in the range $3 \mbox{ MeV}^{-1} \leq \beta \leq 4 \mbox{ MeV}^{-1}$. For each $\beta$, we calculated $G_{\nu}(\tau)$ for a range of values of $\tau$ in steps of $1/32$ MeV$^{-1}$. We chose the asymptotic region in $\tau$ such the linear fits to $\ln G_{\nu}(\tau)$ have a $\chi^2$ per degree of freedom $\sim 1$ or less in all cases considered. We find that a good asymptotic region is  $0.5 \mbox{ MeV}^{-1} \leq \tau \leq 2 \mbox{ MeV}^{-1}$.

Within the asymptotic region, we fit a straight line to $\ln G_{\nu}$ for each possible subset of points in $\tau$ for which $G_{\nu}(\tau)$ has been calculated. The mean and standard deviation of the slopes so obtained are used to estimate $\Delta E_{\rm min}(\ma_{\pm})$ and its statistical error, respectively, at each $\beta$. A weighted average of the results at different values of $\beta$ is then taken.

In a few selected cases, we also performed calculations for larger values of $\beta \,$ (i.e., $\beta> 4$ MeV$^{-1})$, and found the corresponding values of $\Delta E_{\rm min}(\ma_{\pm})$ to be consistent with those obtained in the region $3 \mbox{ MeV}^{-1} \leq \beta \leq 4 \mbox{ MeV}^{-1}$. This indicates that for the model spaces and particle numbers considered, the above chosen values of $\beta$ are sufficiently large to isolate the ground state of the corresponding even-even nucleus.

For a given odd system (an odd-even nucleus) there are two neighboring even systems (even-even nuclei), and our method can be used by starting from either of the even systems. Unless noted otherwise, the results we report here are the average of both of these calculations.

To test the validity and accuracy of our method, we performed calculations in the $sd$ shell using a schematic good-sign Hamiltonian. In all cases, our results deviated no more than $0.1\%$ from the exact ground-state energies,
obtained by diagonalizing the Hamiltonian with the {\tt OXBASH} code~\cite{bro88}. For example for $^{29}$Si we found a ground-state energy of $-133.98 \pm 0.04$  MeV compared with the exact result of $-133.95$ MeV. Our method also reproduced correctly the ground-state spin in all cases.

\begin{figure}[tbp]
\includegraphics[width=0.9\columnwidth]{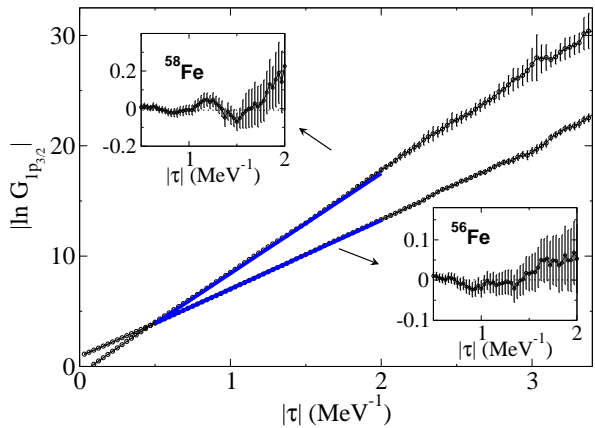}
\caption{The absolute value of logarithm of the Green's function (\ref{green}) for the neutron orbital $\nu=1p_{3/2}$ in $^{56}$Fe (lower curve, $\tau > 0$) and
$^{58}$Fe (upper curve, $\tau \leq 0$)
at $\beta=4$ MeV$^{-1}$.   The solid blue lines are linear fits for $0.5$ MeV$^{-1}\leq |\tau|\leq 2$ MeV$^{-1}$. The insets show the deviations from these linear fits.}
\label{figgffpg}
\end{figure}

We applied our method to nuclei in the ($pf + g_{9/2}$) shell, using the isospin-conserving Hamiltonian of Ref.~\cite{nak97}. Typical results are demonstrated in Fig.~\ref{figgffpg}, in which the absolute value of the logarithm of the Green's functions for the neutron orbital $\nu=1p_{3/2}$ in $^{56}$Fe ($\tau > 0$) and in $^{58}$Fe ($\tau \leq 0$) are plotted versus $|\tau|$ for $\beta=4$ MeV$^{-1}$. The linear fits (solid lines) were used in the calculation of the ground-state energy of $^{57}$Fe. The deviations from the linear fits are shown in the insets of Fig.~\ref{figgffpg}.

\begin{figure}
\includegraphics[width=0.9\columnwidth]{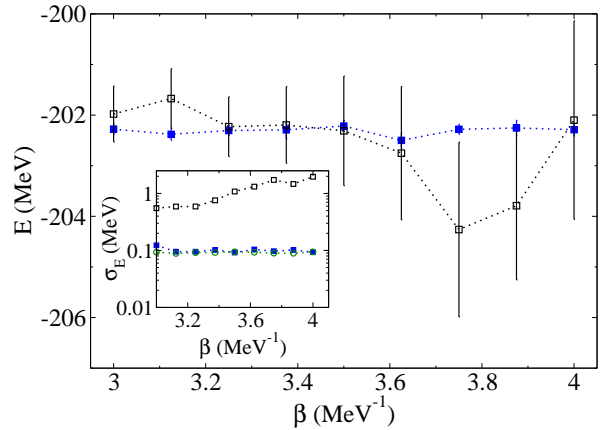}
\caption{The energy of the $^{57}$Fe nucleus calculated from  the present method and direct SMMC are shown by solid and open squares, respectively. The error bars describe the statistical errors.  Inset: the statistical errors for the energy of $^{57}$Fe in the present method (solid squares) and in direct SMMC calculations (open squares) are shown on a logarithmic scale. The statistical errors for the energy of $^{56}$Fe using the same Hamiltonian are shown by open circles.}
\label{figerr}
\end{figure}
A direct application of the SMMC method to the odd-particle systems suffers from a sign problem which leads to very large statistical errors at low temperatures. In contrast, the  method presented here does not have such problem. This is illustrated in Fig.~\ref{figerr} where we compare the energy and its statistical error for the $^{57}$Fe nucleus in the present method (using the neutron Green's functions of $^{56}$Fe) with the results obtained from the direct method. The errors in the present method remain roughly constant with $\beta$.  At $\beta =3$ MeV$^{-1}$ the statistical error in the direct method is about 5 times larger than the present method while at $\beta=4$ MeV$^{-1}$ it is about 20 times larger. The inset shows the statistical errors on a logarithmic scale. For comparison we have also included the statistical error in the energy of the even-even nucleus $^{56}$Fe using the same Hamiltonian.

We applied our Green's function method for families of odd-neutron isotopes: $^{47-49}$Ti, $^{51-57}$Cr, $^{53-61}$Fe, $^{59-65}$Ni, $^{63-67}$Zn and $^{71-73}$Ge. The ground-state spins we determine are in agreement with experimental values in all cases except for $^{47}$Ti, $^{57}$Fe and $^{63}$Ni. The anomalous ground-state spin of $^{57}$Fe from the shell model perspective is well documented in the literature~\cite{ham62}.

\begin{figure}[tbp]
\includegraphics[width=\columnwidth]{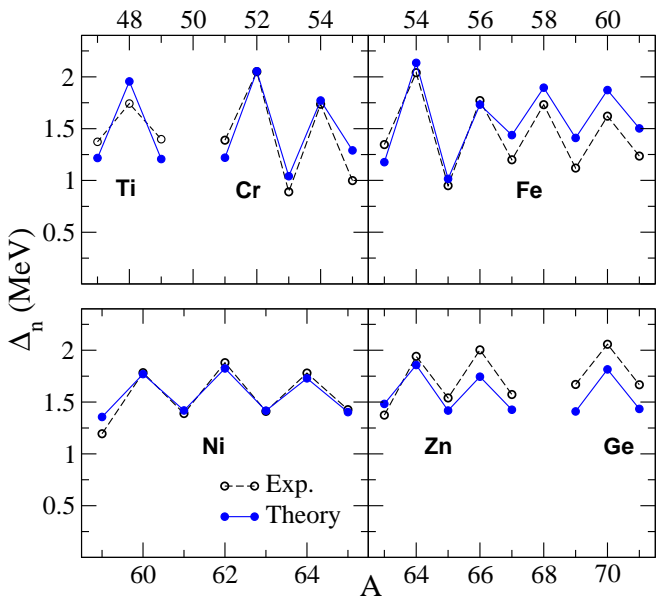}
\caption{Neutron pairing gaps $\Delta_n$ as a function of mass number $A$ in $fp+g_{9/2}$-shell nuclei. The gaps calculated with the present Green's function method (solid circles connected by solid lines) are compared with the experimental gaps (open circles connected by dashed lines). The theoretical statistical errors are smaller than the size of the symbols.}
\label{figpg}
\end{figure}

In our method we extract directly the odd-even ground state energy differences, and therefore this method is particularly suitable for accurate calculations of pairing gaps (i.e., odd-even staggering of masses).

When extracting an odd-even ground-state energy difference such as $\Delta E_{\rm min}(\ma_+)$  we use the Hamiltonian of the  $\ma_+$ nucleus for both the $\ma_+$ and $\ma$ nuclei. Since the $fp+g_{9/2}$-shell Hamiltonian we use is nucleus-dependent~\cite{nak97}, it is necessary to correct the ground-state energy of the $\ma$ nucleus. As the latter is an even-even nucleus, this correction can be found in direct SMMC calculations for the $\ma$ nucleus. However, this correction can also be estimated as follows. The dependence of the interaction on the nucleus is rather weak; the strengths of the multipole-multipole interactions depend weakly on the mass number $A$ ($\propto A^{-1/3}$)
and the monopole pairing strength is constant through the shell.  The largest variation among neighboring nuclei is that of the single-particle energies $\ve_\mu (\ma)$ of the orbitals $\mu$. Correcting for this variation, the neutron separation energy for the $\ma_{+}$ nucleus is given by
\be\label{n-separation}
S_n(\ma_{+}) = - \Delta E_{\rm min} (\ma_{+}) + \sum_{\mu} [\ve_\mu(\ma) - \ve_\mu(\ma_{+})] \lb n_\mu \rb_{\ma} \;,
\ee
 where $\lb n_\mu \rb_{\ma}$ are the average occupation numbers for the $\ma$ nucleus using the Hamiltonian for the $\ma_+$ nucleus. The second term on the r.h.s. of (\ref{n-separation})  approximates the difference between
the ground-state energies of the $\ma$ nucleus when calculated using the
respective Hamiltonians for the $\ma$ and $\ma_+$ nuclei. We verified (in $sd$-shell nuclei) that  this approximation is highly accurate and well within a typical statistical error. In our calculations we used (\ref{n-separation}) since the resulting statistical error is much smaller than the statistical error of direct SMMC calculations.

The neutron separation energy for the $\ma$ nucleus is given by a similar expression. The neutron pairing gaps can then be calculated from the differences of separation energies $\Delta_n(\ma)=(-)^N[S_n(\ma_{+})-S_n(\ma)]/2$, where $\ma$ can now be either an even-even or an odd-even nucleus.

Our calculated pairing gaps are shown in Fig.~\ref{figpg} (solid circles), where they are compared with the experimental pairing gaps (open circles) as determined from odd-even staggering of binding energies. Our results agree quite well with the experimental values; in most cases the deviation of the theoretical pairing gaps from their experimental counterparts is less than $ 15\%$. Systematic deviations are observed for the iron isotopes above $A=59$ and for the germanium isotopes. For the germanium isotopes the size of the model space might be insufficient, while the deviation for the iron isotopes indicates the necessity to refine our isospin-conserving Hamiltonian.

\emph{Conclusion.} We have described a practical method that circumvents a sign problem for calculating the ground-state energy of odd-particle systems in the shell model Monte Carlo approach. We have demonstrated the usefulness of the method by calculating pairing gaps of nuclei in the$fp+g_{9/2}$ shell. This method can also be applied to other finite-size many body systems such as trapped cold atoms. In principle this method can be used more generally to calculate the lowest energy state for a given spin. However, when such a state is an excited state, the statistical errors are larger and it is more difficult to identify the asymptotic regime.

\emph{Acknowledgements.} This work was supported in part by the U.S. Department of Energy Grant No. DE-FG02-91ER40608. Computational cycles were provided by the High Performance Computing Center at Yale University.

\end{document}